\documentclass[12pt]{elsarticle}

\usepackage{amssymb}

\usepackage{hyperref}
\usepackage{tabularx}
\usepackage{multirow}
\usepackage{float}
\usepackage{amsthm}
\usepackage{makecell}
\usepackage{color,soul}
\newtheorem{definition}{Definition}
\newcolumntype{P}[1]{>{\centering\arraybackslash}p{#1}}
\journal{block\-chain: Research and Applications}

\begin{document}

\begin{frontmatter}



\title{Block\-chain software patterns for the design of decentralized applications: A systematic literature review}


\author[first]{Nicolas Six, Nicolas Herbaut, Camille Salinesi}

\affiliation[first]{organization={Centre de Recherche en Informatique, Universit\'e Paris 1 Panthéon-Sorbonne}, country={France}}

\begin{abstract}

A software pattern is a reusable solution to address a commonly occurring problem within a given context when designing software. 
Using patterns is a common practice for software architects to ensure software quality.
Many pattern collections have been proposed for a large number of application domains. 
However, because of the technology's recentness, there are only a few available collections with a lack of extensive testing in industrial block\-chain applications. 
It is also difficult for software architects to adequately apply block\-chain patterns in their applications, as it requires deep knowledge of block\-chain technology. 
Through a systematic literature review, this paper has identified 120 unique block\-chain-related patterns and proposes a pattern taxonomy composed of multiple categories, built from the extracted pattern collection.
The purpose of this collection is to map, classify, and describe all available patterns across the literature to help readers make adequate decisions regarding block\-chain pattern selection.
This study also shows potential applications of those patterns and identifies the relationships between block\-chain patterns and other non-block\-chain software patterns. 

\end{abstract}




\end{frontmatter}


\section{Introduction}
\label{introduction}

Block\-chain technology is a distributed ledger constituted of blocks, supported by a network of peers that each owns a copy of it. 
Each node follows the same protocol and uses a consensus algorithm to keep its copy consistent with others' and manage the addition of new blocks into it.
Users can interact with nodes to append transactions, but modification and deletion of them are theoretically impossible.
While the first generation of block\-chains was only focusing on cryptocurrency transactions between users (e.g., Bitcoin \cite{nakamoto2008bitcoin}), some of them now support smart contracts (e.g., Ethereum \cite{buterin2013ethereum}).
A smart contract is a decentralized program that can be executed on-chain, through nodes. 
Users can deploy and interact with smart contracts using transactions.
Block\-chain is fully decentralized by nature, where no third party is in charge of the network functioning.
Block\-chain data are also immutable and tamper-proof, as nobody can alter a block after its creation and addition, into a block\-chain.
Thanks to these properties, block\-chain-based applications can be trusted, as nobody can tamper with the correct execution of a smart contract\footnote{This is only valid if the smart contract is well designed to prevent execution flaws and security issues.}.
Also, it is possible to retrace the state change history of a block\-chain.
Thus, smart contract state changes can also be replayed for the complete traceability of decentralized applications (dApps).

In recent years, block\-chain has been growing rapidly from a niche technology used by a few people as a promising solution for many sectors, due to its unique properties that empower the design of innovative software architectures and systems \cite{zeadally2019blockchain}.
First, due to the native support of cryptocurrencies, block\-chain enables the creation or the improvement of use cases in the financial domain that was difficult to leverage using existing technologies.
For example, currency exchange through banks can be an expensive process for a consumer, Automated Market Makers (AMM) allow the swap from one cryptocurrency to another without any intermediate using liquidity pools of cryptocurrencies and a smart contract to perform the swap \cite{pourpouneh2020automated}. 
Regarding insurance, block\-chain can be used to automate the claiming process in case of an accident. Such a process would take many days or weeks with traditional insurance systems \cite{oham2018b}.
Block\-chain also has many applications in nonfinancial domains, due to its capacity to operate without any third party and enable trust with the usage of decentralized applications.
For instance, block\-chain can be the platform in an inter-organizational business process, to monitor organizations' actions and data, or to allow the business process execution directly on-chain \cite{di2019blockchain}.
In this context, participants can trust the information stored by the block\-chain, and operations performed in smart contracts cannot be tampered with.
This layer of decentralized automation and trust is also used in other applications, such as smart grids \cite{agung2020blockchain}, as block\-chain can connect thousands of individuals to enable a market for energy exchange between users, or healthcare for medical records sharing.
As block\-chain is increasingly considered in many use cases, many companies started to show interest in block\-chain and build new applications.
According to the Deloitte's 2020 Global block\-chain Survey\footnote{\url{https://www2.deloitte.com/ie/en/pages/technology/articles/Global_blockchain_survey.html}}, 55\% of the 1488 surveyed companies across the world considers block\-chain as one of their top-five strategic priorities.
However, even if more and more companies start block\-chain projects, only a few are successful.
Organizational, legal, and technical issues still hinder the adoption of block\-chain.

Regarding technical issues, it is indeed tedious for software architects to include block\-chain in new or existing applications. 
First, they have to consider the liabilities of block\-chain technology: depending on its type, high latency and bad throughput can be observed, as the network needs to reach a consensus to process a new block and its transactions.
Data immutability, a quality in some cases, can also be a burden: it can interfere with data protection regulation such as GDPR \cite{voigt2017eu}, or hinder the upgradeability and the modifiability of smart contracts.
Second, they have to be careful about the design of smart contracts.
Because of the non-repudiation aspect of block\-chain, every vulnerability in a smart contract can be a major security threat that leads to a loss of funds or erratic behavior.
Design flaws can also generate extra functioning costs, augmented by the fact that public block\-chains often require the usage of cryptocurrencies to pay for smart contract execution.

In software engineering, software patterns are used to tackle such design issues. 
Specified as repeatable solutions for commonly recurring problems, they ensure that the final design of the software fulfills expected requirements.
Therefore, an adequate selection of software patterns can lead to a well-designed architecture.
For block\-chain-based applications, this is difficult to achieve, and the design of block\-chain-based software architecture is one of the main challenges in the block\-chain-Oriented Software Engineering (BOSE) field \cite{porru2017blockchain}.
Unfortunately, there are only a few block\-chain-based patterns available, scattered throughout the literature and industrial websites.
The software architect can also struggle with the selection of adequate patterns as it requires extensive knowledge of block\-chain for pattern selection.
For other technological fields, systematic literature reviews of patterns have been proposed to collect existing patterns across the literature (see Section \ref{related}).
To the best of our knowledge, such a study does not exist yet for block\-chain-based patterns, despite the benefit of it for the state of the art of block\-chain-based design patterns.

Throughout a systematic literature review, this paper tries to fill the gap by collecting all available block\-chain-based patterns and classifying them into comprehensive categories that help to fetch adequate block\-chain patterns.
A detailed description of the most frequently mentioned patterns is also given through a pattern format.
This paper also investigates the link between block\-chain patterns and the connections between them and non-block\-chain software patterns.
A description of the application domains of those patterns is also given.
In parallel, this paper also proposes some future research directions for block\-chain-based patterns, as gaps were identified throughout the study.

The main contribution of this study is a uniform collection of 160 block\-chain-based software patterns from 20 different academic sources, reduced to 120 unique patterns.
A design pattern taxonomy is also proposed to classify current and upcoming block\-chain-based patterns in comprehensive categories.
Those contributions have been stored on GitHub\footnote{https://github.com/harmonica-project/blockchain-patterns-collection}, to allow their reusability by others.
This work aims to facilitate navigating across existing patterns for readers, through a comprehensive classification, a short rationale on many patterns, and a mapping between identified patterns, their variants, and the associated papers.

The paper is structured as follows: Section \ref{background} introduces some background on block\-chain technology and software patterns, then Section \ref{methodology} describes the review process employed and the results obtained, discussed in Section \ref{discussion}.
Threats of validity are discussed in Section \ref{threats}, Section \ref{related} introduces related works, and section \ref{conclusion} concludes the paper with planned future works.

\section{Background}
\label{background}

\subsection{Block\-chain}

The first implementation of block\-chain technology has been proposed in 2008 when Satoshi Nakamoto released a whitepaper on Bitcoin, a decentralized cryptocurrency \cite{nakamoto2008bitcoin}.
He combined multiple previously existing technologies, such as asymmetric encryption \cite{simmons1979symmetric}, Merkle tree structures \cite{merkle1989certified}, consensus methods, and Hashcash, a cryptographic algorithm where computing the proof is difficult and verifying it is a simple task \cite{back2002hashcash}. 
This combination has defined the foundation of block\-chain technology and its associated network.

According to \cite{belotti2019vademecum}, a possible definition of block\-chain is the following (\ref{def-bc}):

\begin{definition}
\label{def-bc}
A block\-chain is an immutable read-only data structure, where new entries (blocks) get appended onto the end of the ledger by linkage to the previous block’s ‘hash’ identifier.
\end{definition}

Usually, blocks contain transactions, where their type depends on the block\-chain usage: for Bitcoin, transactions represent an exchange of cryptocurrency between users.
The block\-chain as a data structure is maintained by a network of peers.
Each member of the network owns a copy of the block\-chain.
They communicate using the same protocol to maintain their copy up to date.
To do that, each block\-chain protocol comes with its consensus algorithm, a mechanism to reach an agreement among participating nodes on the transaction order.
To mention a few of them, the Proof-of-Work algorithm is based on a mathematical challenge that nodes have to solve for appending a block to the block\-chain \cite{gervais2016security}.
If so, they can share the block with others and start searching for a solution for the next block.
Using the Proof-of-Stake algorithm, participants must put collateral at stake to be entitled to create and share blocks\footnote{The term minting is often employed in the context of PoS-based block\-chains for this operation.}.
The size of the collateral determines the share of the blocks it has the right to create \cite{saleh2021blockchain}.
Misbehaving nodes are punished by taking out their stakes.
Another notable algorithm is the Practical Byzantine Fault-Tolerant (PBFT) algorithm, that tolerates Byzantine faults (e.g., dysfunctional or malicious nodes) in the network \cite{sukhwani2017performance}.
A leader, once elected, is responsible for broadcasting new transactions from clients to backup nodes that verify the transaction, execute the required operations, and then propagate the transaction.
If enough backup nodes agree on the same result, the transaction is appended to the block\-chain.

For the first two consensus algorithms, participants must put at stake something with real-world value: either computing power or cryptocurrencies.
However, for the third one, there is nothing at stake: the only solution for a secure network is to know the participants and exclude them in case of misbehaving.  
Depending on the consensus algorithm used, block\-chain networks can either allow anybody to join and participate or require approval from others to join, called respectively public and private block\-chains.
Selecting the right block\-chain for a given context is a tough choice: public block\-chains are more decentralized than private ones in general, as anybody can join and participate, but might suffer from bad performance due to heavy consensus algorithms, where private block\-chains are more efficient but often controlled by a group of organizations.
For example, Bitcoin can only process 6 transactions per second using a Proof-of-Work algorithm, whereas Hyperledger Fabric with PBFT can process hundreds of transactions per second.

Block\-chain can be used to transact cryptocurrencies, but also leverage decentralized applications, so-called smart contracts.
The first proposal of block\-chain smart contracts traced back to 2015 with Ethereum \cite{buterin2013ethereum}, and generalized to many block\-chains since then.
A smart contract is a decentralized on-chain program that can be instantiated and requested through transactions. 
Executing a smart contract function follows the same process as adding a transaction into the block\-chain: the function is performed by the requested node, and the result is shared with the other nodes that will also verify the correctness of the function execution.
When building decentralized applications, we can differentiate its off-chain part from its on-chain part.
The on-chain part is usually constituted by smart contracts, and the off-chain part is composed of components that are not part of the network but might interact with it.
The distinction is important as it constitutes a separation between patterns in the taxonomy presented in Section \ref{rq1}.

Through its specific behavior, block\-chain technology has many interesting properties:

\begin{itemize}
    \item Decentralization by nature - no one is in charge of the whole network. By extension, smart contract-based apps are also decentralized, as no third party is responsible for executing its functions and returning the result to others.
    \item Transparency - every network participant can dive into the content of the block\-chain, either transactions or smart contracts data.
    \item Tamper-proofing and immutability - it is impossible to modify the content of a block after its addition. It would be detected by others because the hash of the block would change and mismatch the block hash already stored in the next block.
\end{itemize}

However, block\-chain qualities can also be liabilities, depending on the context.
The transparency and immutability of a block\-chain can put personal or confidential data at risk.
Even encrypted, it is unsure that data is safe, because of potential advances in data decryption or key leakage.
Immutability also means it is impossible to reverse transactions, even if they are harmful.
As an example, a vulnerability exploited in TheDAO smart contract has led to a loss of 12 million \$USD in Ether, the network cryptocurrency\footnote{\url{https://www.coindesk.com/understanding-dao-hack-journalists}}. 
Finally, the poor performance of block\-chains can also be a burden, when low latency or high throughput is expected.
Thus, any company that wants to use block\-chains in their applications must carefully assess the implications, as this is not always the best solution.
Software patterns can help to lower the impact of block\-chain liabilities on the final design, to guide the design of block\-chain applications through repeatable solutions, or to ensure that block\-chain qualities are kept intact in the final design.
However, there is still a lack of a wide structured collection of software patterns for block\-chain.
This paper will address several research questions aiming towards this goal.

\subsection{Software pattern}
\label{patternsbg}
In the software engineering field, patterns are strong assets for engineers and architects to design robust and well-designed applications.
The principle of patterns was first proposed by Christopher Alexander in the construction field, as he proposed to document architecture designs in a way that documentation can be reused for other buildings \cite{alexander1979timeless}.
In one of his books \cite{alexander1977pattern}, he proposes a definition of patterns, commonly reused later by other researchers, that is the following (\ref{def-pattern}):

\begin{definition}
\label{def-pattern}
Each pattern describes a problem that occurs over and over again in our environment, and then describes the core of the solution to that problem, in such a way that you can use this solution a million times over, without ever doing it the same way twice.
\end{definition}

In the software engineering field, patterns appeared later in 1987 where Cunningham et al. decided to apply the pattern approach to guide developers using Smalltalk, an object-oriented language \cite{beck1987using}.
Later on, 4 researchers (commonly called the GoF - Gang of Four) released a book that defines a collection of design patterns for the development of object-oriented applications \cite{gamma1995elements}.
Since then, many researchers have proposed software patterns for many uses cases, such as microservices \cite{taibi2018architectural} and Internet-of-Things (IoT) \cite{qanbari2016iot}.

They can be grouped into three categories: architectural patterns that define, at the highest level of abstraction, the general structure of the application (elements, connections), design patterns that define a way to organize modules, classes, or components to solve a problem, and idioms, a solution to a language-related problem at the code level.
Using patterns in an application brings many advantages.
First, as existing patterns are often extensively tested and applied by others, they can be reused in a new design as the best solution possible for a given case.
They also define a common language among developers, as software patterns are defined with a meaningful name.
However, their application must not always be systematic: applying the wrong pattern to a certain design can be more harmful than helpful.
They might also increase the complexity of software.
As an example, the \textit{Proxy} pattern, that helps to control the access to an object is unnecessary if the object in question is not sensitive and only accessed by one other object.

To be easily reused, software patterns are often expressed using a pattern template.
The two most commonly used pattern templates are the form proposed by the GoF (GoF pattern format), and the Alexandrian form, by Christopher Alexander \cite{tevsanovic2005pattern}.
In both approaches, a pattern is described by an expressive \textbf{Name}, the \textbf{Context} it is applicable to, and a recurring \textbf{Problem}. 
The Alexandrian form is also constituted by the following: the \textbf{Solution} to describe the pattern, the \textbf{Forces} where the pattern has an impact on, \textbf{Examples} of application, the \textbf{Resulting context}, a \textbf{Rationale} on deep or complex aspects of the patterns, \textbf{Related pattern} and \textbf{Known uses}.
The GoF format contains other types of information: an optional \textbf{Classification} of the pattern among others, a \textbf{Known as} field in case the pattern also exists with different names, the \textbf{Motivation} to introduce an example of scenario the pattern can address, \textbf{Applicability} to describe situations where the pattern can be applied, \textbf{Participants} (eg. classes and objects) and the \textbf{Collaboration} that links them to carry out their responsibilities, the \textbf{Structure} of the pattern, the \textbf{Consequences} of using it on the software, an \textbf{Implementation} part to describe code samples and key technical aspects to consider, \textbf{Known uses} and \textbf{Related pattern}.
The process of writing patterns from existing knowledge is also a subject of research in the pattern community.
For example, \cite{meszaros1997g} proposes a pattern language for pattern writing, thus using patterns to address commonly occurring problems when writing patterns.
\cite{harrison1999language} presents advice for shepherding, a method used in the pattern community to improve the quality of patterns by having an experienced pattern writer review patterns from others.
The patterns format as well as the methodologies to write patterns are very useful to construct patterns in a comprehensive and informative way, as it can be difficult to formalize a pattern even with expertise in the domain associated with the pattern to write.
In this paper, we have chosen to use a reduced pattern format, as there is no uniform notation in collected papers to express patterns with the same amount of detail as the Alexandrian or GoF pattern format.
Hence, patterns were collected during the literature review as a \textbf{Context and problem}, a \textbf{Solution}, and \textbf{Examples}.
Using this pattern format, a uniformized collection of patterns was constituted.

\section{Review Process}
\label{methodology}

As it is important to follow a robust methodology to perform a high-quality literature review, this paper follows Kitchenham et al. \cite{Kitchenham07guidelinesfor} guidelines to conduct a Systematic Literature Review (SLR).
This task was divided into three main stages as follows:
\begin{enumerate}
    \item Planning: during this phase, the research questions, as well as the goals of the SLR, are elicited. Also, the literature databases that will serve for the retrieval of papers are selected, and inclusion/exclusion criteria are given.
    \item Conducting: the SLR is conducted, following the plan designed earlier. Studies are extracted then filtered, and the remaining papers are read. An analytic framework is used to extract the necessary data to answer the research questions. 
    \item Reporting: results of the SLR are factually given, as well as a quality assessment of the extracted studies. Then, they are discussed in their own section. 
\end{enumerate}

\subsection{Review Planning}
\label{planning}

The first step in the planning of the systematic literature review is the formalization of sound research questions.
Those questions have to be designed considering that the answers must address the research goals of this study.
The main purpose of this study is the design of a comprehensive and uniform collection of block\-chain software patterns extracted from the existing literature.
However, collecting the patterns in bulk is not enough to allow their reusability and usability; thus a classification scheme must be proposed along.
To further refine the quality of extracted patterns, we can also consider the context of those patterns: their relation with existing non-block\-chain patterns (e.g. GoF patterns \cite{gamma1995elements}) or their links with specific technologies or domains. 
Indeed, we found several patterns that cannot be separated from their domain or their technology.
As an example, the \textit{Limit modifiers} pattern is directly bound to the modifier keyword in the Solidity language, thus non-applicable to block\-chains that do not support it.
These aspects must be addressed in research questions.
Finally, the results of the systematic literature review can be used to highlight several research gaps in the block\-chain software pattern literature for further exploration.
From the different considerations of this study, the following research questions have been formulated:

\begin{itemize}
    \item \textbf{RQ1}: What taxonomy can be built from existing literature on block\-chain-based patterns?
    \item \textbf{RQ2}: What are the existing block\-chain-based patterns and their different categories?
    \item \textbf{RQ3}: What are the most frequently mentioned patterns and their variants across identified patterns?
    \item \textbf{RQ4}: Are some of the patterns equivalent to existing software patterns?
    \item \textbf{RQ5}: What are the applications of the literature patterns?
    \item \textbf{RQ6}: What are the current gaps in research on block\-chain-based patterns?
\end{itemize}

Three library databases have been selected to extract relevant studies: IEEE Xplore, ACM Digital Library, and Scopus.
Snowballing from selected papers is also considered as a data source, as it might help to include other relevant papers.
To query the databases of papers, a search query has to be designed.
We have chosen to use the Quasi-Gold Standard (QGS) method to select the words composing the query.
The QGS method consists in selecting a set of studies that must appear in the results of the query, then designing the query around the terms employed in those papers.
Thus, 5 studies have been selected to compose this corpus of studies \cite{13,9,27,35,14}.
From that, the following query has been constituted:

\begin{quote}
\textit{(block\-chain OR block\-chain-based OR "smart contract*") AND ("idiom*" OR ("architectural pattern*" OR "design pattern*" OR "block\-chain pattern*" OR "block\-chain-based pattern*"))}
\end{quote}

We decided to include only the studies that have those terms in their title, abstract, or keywords, to improve the precision of the query. 
To prepare for the filtering phase of the SLR, inclusion and exclusion criteria have been defined.
They provide systematic guidelines to include or exclude papers during the filtering phase, where papers are selected for further reading.
Table \ref{criteria} provides the chosen inclusion and exclusion criteria.

\begin{table}[h]
\begin{tabularx}{\textwidth}{|X|X|}
\hline
\multicolumn{1}{|c|}{\bfseries Inclusion criteria} & \multicolumn{1}{c|}{\bfseries Exclusion criteria} \\ \hline
- Presents one or more block\-chain-based patterns. & - The paper is a duplicate of other studies.  \\
- The paper is described as presenting block\-chain-based patterns in other accepted studies. & - The paper lies outside the software engineering and block\-chain domains.  \\
 & - Full text is not accessible.   \\
 & - The paper is not written in English. \\
 & - The paper has not been peer-reviewed. \\
\hline
\end{tabularx}
\caption{Inclusion and exclusion criteria.}
\label{criteria}
\end{table}

Finally, a set of questions have been prepared to assess the quality of the extracted patterns:
\label{qqlabels}

\begin{itemize}
    \item \label{qq1} \textbf{QQ1}: Does the paper clearly present the pattern solutions, problems, and contexts?
    \item \textbf{QQ2}: Does the paper reference existing solutions using presented patterns?
    \item \textbf{QQ3}: Does the paper use a standard pattern presentation form (e.g., GoF/Alexandrian templates \cite{tevsanovic2005pattern}, described in Subsection \ref{patternsbg})?
\end{itemize}

For each question, an answer can be given among the following options: "Yes", "Partially", and "No".
Knowing the answer for a paper can help to assess the quality of the patterns introduced in it, where knowing the answer to all the papers assesses the quality of the collection derived from this literature study. 
To guarantee the quality of the extracted patterns, we decided to only keep papers where the answer to the first question is at least "Partially".
Indeed, it is difficult to extract a clear pattern where there is no description of the solution and the problem it addresses in a specific context.

\subsection{Review Execution}

\begin{figure}[h!]
    \centering
    \includegraphics[scale=1]{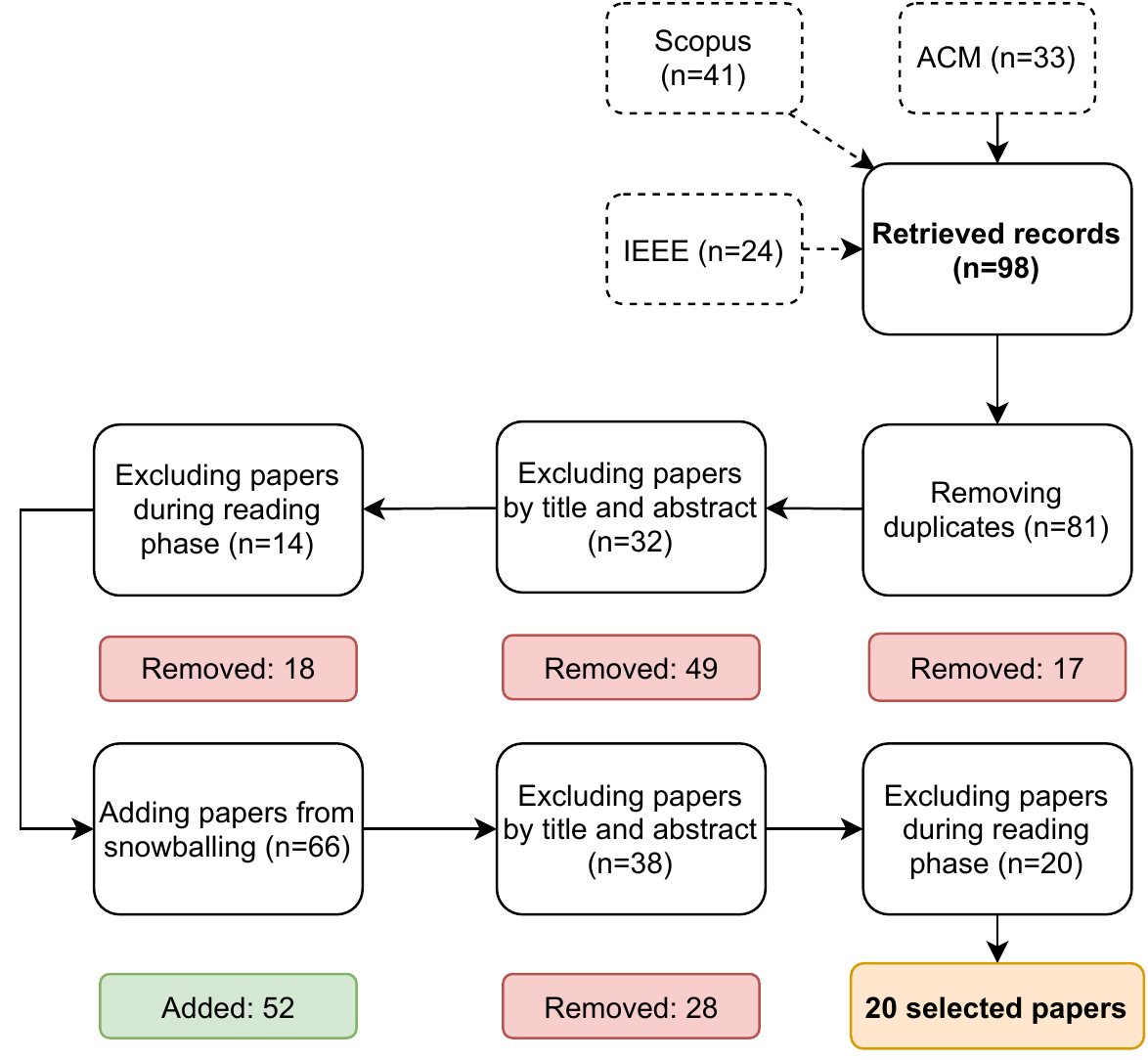}
    \caption{Review process scheme.}
    \label{review}
\end{figure}

Figure \ref{review} gives a graphical overview of the review protocol, where for each step the number of remaining or excluded papers is displayed.
98 papers have been retrieved using the query over the three selected databases.
As Scopus indexes papers from many other libraries, 17 duplicates were found and removed.
Then, papers have been filtered on their title, abstract, and keywords based on the inclusion/exclusion criteria defined in the review planning (\ref{planning}).
32 papers were kept from this first filtering.
18 additional papers were filtered out during the reading phase, for several reasons. 
First, some of them were not fitting our inclusion/exclusion criteria, as they were not presenting any design patterns in their studies.
Also, the presentation of software patterns in several papers was not clear enough for data extraction (\hyperref[qq1]{QQ1}).
Lastly, some papers were excluded as they were merely presenting patterns without proposing any enhancement.
During the reading phase, papers that were mentioned by others to introduce block\-chain-based patterns were added to the corpus of papers.

In addition, backward and forward snowballing was done for each paper to complete the corpus of studies.
Regularly performed during systematic literature reviews, backward and forward snowballing respectively aims to analyze the citations of selected papers and other papers that cited selected papers to find new relevant papers.
This has led to the addition of 52 papers from snowballing, where 46 of them were filtered out.
The result is the addition of 6 new studies into the final corpus, that were exclusively found during backward snowballing.
Forward snowballing hasn’t yielded any new study into the corpus.
Note that, contrary to forward snowballing and regular inclusion of papers through performed queries, non-peer-reviewed papers were not excluded during backward snowballing. 
This decision has been made as they can be considered relevant, as selected papers citing them were peer-reviewed themselves.

\subsection{Taxonomy construction}
\label{taxo-const}
In parallel with the review process, the taxonomy was built using newly acquired knowledge.
To achieve such a task, a taxonomy development methodology was used \cite{nickerson2013method}.
The methodology proposed by Nickerson et al. first describes what a taxonomy is and the associated problems for taxonomy development.
Then, it gives a method for taxonomy development that satisfies the problems mentioned before, adaptable for many contexts.

According to \cite{nickerson2013method}, a possible definition of a taxonomy is the following (\ref{def-taxo}):

\begin{definition}
\label{def-taxo}
A taxonomy is a set of dimensions each consisting of mutually exclusive and collectively exhaustive characteristics such as each object under consideration has one and only one characteristic for each dimension.
\end{definition}

An important attribute, as stated by the definition, is that no object can have two different characteristics in a dimension.
Also, a taxonomy is not meant to be perfect and can change over time, but they have to fulfill qualitative attributes to be usable.
A taxonomy must be concise: too many dimensions can lead to difficulties in applying the taxonomy.
It must also be robust, containing enough clear dimensions and characteristics to differentiate objects contained inside, and comprehensive, that is the capability to classify as known objects within the domain under consideration.
Finally, a taxonomy must be extensible to adapt to the needs of users and enable the inclusion of new objects, and explanatory to provide information on the nature of the objects under study.
Those qualities are particularly important for the construction of our taxonomy: the conciseness and the robustness of the taxonomy will help the reader to navigate in the different categories available to pick relevant patterns (i.e., the knowledge domain), and the extensibility will allow the taxonomy to grow over future studies on block\-chain patterns.

The first step of the taxonomy construction, as presented in \cite{nickerson2013method}, is to define meta-characteristics.
This gives a basis for identifying the other characteristics of the taxonomy.
In this taxonomy, the two meta-characteristic "On-chain pattern" and "On/off-chain interaction pattern" have been chosen.
As this study focuses on design aspects, we found it relevant to order patterns depending on their position regarding the block\-chain: in the block\-chain (smart contracts, transaction data), or out of the block\-chain (services that interact with the block\-chain, wallets, ...). 
Then, as building a taxonomy is an iterative process, ending conditions must be determined.
Indeed, as said before, a taxonomy is not meant to be perfect; thus the process must be stopped when a satisfactory taxonomy is built.
It is considered satisfactory when all the qualities of a well-built taxonomy are present.
Additional ending conditions can also be added.
One condition we chose to add is the following: all objects of a representative sample of objects have been examined.
As the patterns are the cornerstone of this study, it is important to examine all of them to construct an accurate taxonomy.
Therefore, this taxonomy construction is empirical-to-conceptual rather than the opposite: from the patterns, categories are drafted and then refined to return an accurate taxonomy.

The next three steps are the construction of the taxonomy itself.
As they are incremental, they must be repeated until ending conditions are met.
To begin, identification of a subset of objects must be done.
In our case, the subset is constituted of all the identified patterns.
The next step is identifying common characteristics and group objects.
To do that, a Natural Language-based algorithm was designed to ingest all the summaries of patterns, lemmatize them, and identify a recurrent suite of words (n-grams).
For bigrams, the most recurrent combination of words were "Smart contract(s)" (54 times), "Data storage" (11 times), "Proxy contract" (6 times), and "Factory object" (6 times).
Other interesting combinations were found: "Outside block\-chain" (5 times),  "Restrict execution" (3 times), and "Critical operation" (3 times).
From those combinations and others, three assumptions can be made: (1) smart contract is a crucial topic in block\-chain-based patterns, (2) many traditional software design patterns were found in patterns summaries. Thus links might exist between the existing knowledge on software patterns and newly designed patterns, (3) some important design aspects are recurrent in pattern summaries.
Existing collection names were also exploited to generate categories.
For example, \cite{20} proposes patterns exclusively dedicated to smart contract gas efficiency.
Such collection gives hints of potential types of categories.
Using those assumptions and our personal knowledge, a first taxonomy has been built.
As not all of the patterns were fitting defined dimensions in the first iteration, two other iterations were performed to construct the version of the taxonomy presented in this study.
We also found during the literature review that the majority of found patterns are design patterns, thus the taxonomy has been recentered from all software patterns to design patterns.
The final version of this taxonomy is presented in the Subsection \ref{rq1} associated with the RQ1.

\subsection{Results}

This section factually presents the results of the systematic literature study.
More details are given when discussing each research question in Section \ref{discussion}.

The final corpus of papers is composed of 20 studies, out of which 6 were added through reference snowballing.
19 of them propose design patterns, whereas only one proposes architectural patterns.
No study that introduced idioms was found.
However, some of the patterns found were more related to idioms than design patterns and categorized as such.
160 patterns were found from these 20 studies, including duplicates.
At first, patterns that were said to come from other studies were also added but filtered out afterward to ensure no pattern is missing from the extracting phase.
After duplicate removal, 120 unique patterns have been found.
104 of them have been classified as design patterns, 3 of them as architectural patterns, and 14 as idioms.
As the links between patterns across papers were collected during the SLR, they have been used to filter a large number of duplicates.
Then, pattern names and summaries/solutions were used to filter out additional patterns.
Precautions have been taken when removing patterns using those fields: close patterns that diverge on tiny aspects were kept as separate patterns.

\begin{figure}[H]
    \centering
    \includegraphics[scale=0.6]{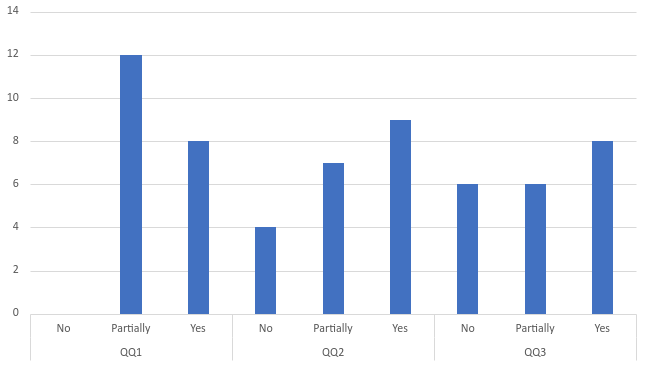}
    \caption{Quality assessment answers distribution (labels detailed in Subsection \ref{qqlabels})}
    \label{qq}
\end{figure}

Regarding the quality assessment performed on accepted papers, Figure \ref{qq} shows the distribution of the answers to each question.

For the first quality question, 8 papers out of 20 clearly introduce patterns, whereas 12 papers might lack details in the pattern detailing.
The second quality question shows that 4 papers do not mention any example of implementation, 7 references one example on average per pattern, and 9 studies reference more than 2 implementation examples.
Finally, the third quality question indicates that 8 papers are using a pattern format to describe their patterns, 6 papers are using a form but lack important sections usually found in pattern formats, and 6 studies do not use any particular format.

\section{Discussion}
\label{discussion}

In this section, each research question is addressed using the results collected throughout the completion of the systematic literature review. 
An answer is given for each question in its respective subsection, and a discussion is done of those results.

\subsection{RQ1: What taxonomy can be built from existing literature on block\-chain-based patterns?}
\label{rq1}

\begin{figure}[h]
    \centering
    \includegraphics[scale=0.67]{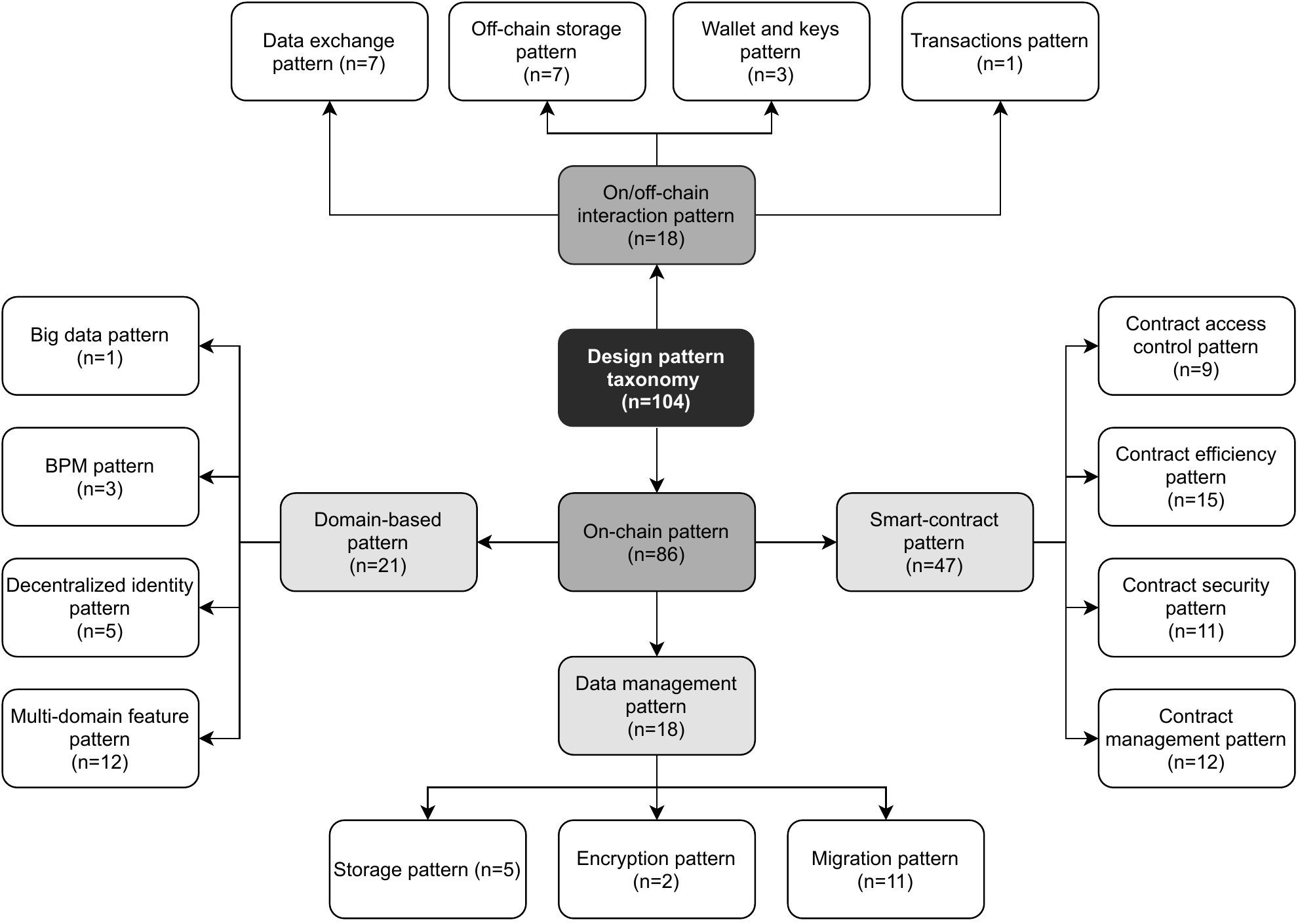}
    \caption{Design pattern taxonomy.}
    \label{scheme}
\end{figure}

In this subsection, a taxonomy of design patterns is presented to classify the design patterns in comprehensive categories that help to decide on what patterns to use for a specific aspect of block\-chain-based application development.
This taxonomy has been built using the methodology from \cite{nickerson2013method}, and its construction is detailed in Subsection \ref{taxo-const}.
Figure \ref{scheme} shows a graphical representation of the proposed design pattern taxonomy.

The taxonomy consists of 2 meta-categories, "On-chain pattern" and "On/off-chain interaction pattern", and 15 different categories.
Intermediate categories were also created to group categories together in the "On-chain pattern" meta-category: "Smart contract pattern", "Data management pattern", and "Domain-based pattern".

The "On/off-chain interaction pattern" category aims to regroup design patterns constituted of off-chain elements that interact with a block\-chain.
This category is key for the development of decentralized applications, as proposed design patterns might bridge off-chain systems and software with on-chain data or smart contracts.
Four subcategories compose this category. The first one, "Data exchange pattern" subcategory, groups patterns that enable communication between on-chain smart contracts and off-chain components.
Indeed, block\-chain cannot request data from outside, thus requiring an external service (i.e. Oracle) to push fresh data inside smart contracts.
The "Data management pattern" subcategory is comprised of design patterns that leverage off-chain data but use block\-chain to guarantee tamper-proofing or trustability of those data.
For instance, hashing a dataset, then storing the hash on-chain to attest later the integrity of the dataset.
The "Wallet and keys pattern" subcategory tackles the management of wallets and keys in the context of a decentralized application.
Finally, "Transactions pattern" subcategory deals with the transaction aspects between off-chain components and the block\-chain, such as transaction confirmation or block inclusion.

In the "Domain-based pattern" intermediate category, on-chain patterns that deal with domain features are regrouped.
Note that this category is meant to be extended with the advances in block\-chain-based patterns for specific domains.
Therefore, three domain-specific categories were created from the knowledge of existing domain-based patterns: "Business Process Management (BPM) pattern" subcategory concerns on-chain business process management (e.g., on-chain activities, ...), the "Big data pattern" subcategory proposes applications of block\-chain for big data, and the "Decentralized identity pattern" subcategory leverage block\-chain to create and manage decentralized identities.
A fourth subcategory, "Multi-domain feature pattern", contains features that do not belong to a particular domain but rather can be used by multiple domains.

The "Smart contract pattern" intermediate category classifies patterns that concern smart contract implementation and management.
As ensuring the security of smart contracts is primordial, the "Contract security pattern" subcategory regroups smart contract patterns that deal with security issues such as reentrancy attacks, overflow attacks, or flawed behavior of smart contracts.
The "Contract efficiency pattern" subcategory essentially deals with patterns that reduce the price of leveraging smart contracts, especially on public block\-chains.
It also contains patterns on other efficiency aspects such as data refreshing, a difficult task with smart contracts as they cannot perform requests on others smart contracts by themselves.
The "Contract access control pattern" subcategory regroups patterns for permission and authorization management for the execution of smart contract functions.
Finally, the "Contract management pattern" subcategory helps with designing the organization of smart contracts together.
For example, having a proxy smart contract that relays the function calls to other contracts.

The last intermediate category, "Data management pattern" deals with patterns for efficient on-chain data management.
It is different from the "Data management pattern" subcategory of "On/off-chain interaction pattern" subcategory as it only concerns data on-chain, located in smart contracts or directly in transactions.
The "Migration pattern" subcategory groups patterns that help with migrating data from one block\-chain to another.
Under "Encryption pattern" are classified patterns for on-chain data encryption, and "Storage pattern" regroups patterns that deal with on-chain data storage.

Through the systematic literature review, the taxonomy has been applied to classify patterns with success.
However, it is meant to be extensible; thus categories might be changed depending on the evolution of the state of the art in block\-chain-based patterns notably with the appearance of new architectural patterns or idioms, not present in this taxonomy due to their scarcity.

This taxonomy is important for an adequate usage of patterns identified in the systematic literature review.
For example, a user willing to implement smart contract security measures in his application to protect it against threats or vulnerabilities will be tempted to search in the "Contract security pattern" subcategory instead of directly searching in the corpus of patterns.
They are also complementary: as each category covers a specific aspect of the design of a block\-chain-based application, they can be combined depending on the user requirements.
For instance, "Contract security" patterns can be used along "Contract efficiency" patterns to improve at the same time the cost efficiency and the security of designed smart contracts.
However, possible conflicts between individual patterns are left outside the scope of this paper, as this information is not present in retrieved papers.

\subsection{RQ2: What are the existing block\-chain-based patterns and their different categories?}
\label{rq2}

This subsection introduces examples of patterns found within the different categories defined by the taxonomy.
For the sake of brevity, not all of the patterns found will be introduced.
The focus will notably be made on patterns observed in multiple studies.
However, the results of this study including the list of all patterns are available on GitHub\footnote{https://github.com/harmonica-project/block\-chain-patterns-collection}.
Note that even if architectural patterns and idioms are outside of the taxonomy, they will still be presented at the end of the subsection.

\subsubsection{On/off-chain Interaction Patterns}

This first category regroups all of the patterns with their components both on and off-chain.
It is divided into four subcategories.
Table \ref{interactionpattern} lists all the patterns contained in this category.

\parindent=0cm

\begin{table}[H]
\begin{tabular}{|P{4cm}|m{8,8cm}|}
 \hline
 \multicolumn{2}{|c|}{\textbf{On/off-chain interaction patterns}} \\
 \hline
 \multicolumn{1}{|c|}{\textbf{\small Subcategory}} & \multicolumn{1}{c|}{\textbf{\small Patterns}}\\
 \hline
    Data exchange pattern & Ticker tape \cite{5} - Oracle \cite{11,13,27,9} - Reverse Oracle \cite{11,13,5} - Pull-based inbound oracle \cite{39} - Push-based inbound oracle \cite{39} - Pull-based outbound oracle \cite{39} - Push-based outbound oracle \cite{39} \\
    \hline
    Data management pattern & State Channel \cite{11,13} - (Off-chain) Contract Registry \cite{11} - Legal and smart contract pair \cite{13} - Off-chain data storage \cite{2,11,13,14,25,36} - Confidential and pseudo-anonymous contract enforcement \cite{23} - Off-chain Signatures \cite{36} - Delegated Computation \cite{36} \\
    \hline
    Wallet and keys pattern & Master \& Sub Key \cite{14} - Hot \& Cold Wallet Storage \cite{14} - Key Sharding \cite{14} \\
    \hline
    Transactions patterns & X-confirmation \cite{13} \\
 \hline
\end{tabular}
\caption{On/off-chain interaction patterns.}
\label{interactionpattern}
\end{table}

The first one is named "Data exchange pattern", to group patterns that enable communication between on-chain smart contracts and off-chain components.
7 patterns were sorted in this subcategory. 
The most frequent pattern is the \textit{Oracle} pattern, introduced or mentioned by 5 different papers \cite{11,13,27,9,5}.
As block\-chain cannot request the external world to retrieve up-to-date information, components named oracles have been designed to listen for block\-chain requests or statuses that indicate some information is needed, then send a transaction to the block\-chain to inject them.
Its opposite has also been proposed: the \textit{Reverse oracle} pattern is applied when off-chain components need block\-chain data to work, so they listen for specific state changes and react accordingly \cite{5, 13, 11, 20}.
Another study proposed more detailed variants of those patterns, as they differentiate the data flow direction (as the \textit{Oracle} and \textit{Reverse Oracle}), as well as if data are pushed out of the data source or pulled from an active component. 

The second subcategory groups 7 patterns that manage and store data off-chain while using block\-chain as an additional layer of trust.
A commonly proposed pattern under many names is the \textit{Off-chain data storage} pattern \cite{2, 11, 13, 14, 25, 36}.
It consists of storing large amounts of data off-chain, then producing a hash of the data and saving it on-chain. 
Therefore, it is far cheaper to leverage while having a possibility to check the integrity of stored data using the hash on-chain.
This pattern is presented in detail in a dedicated part of Subsection \ref{p-offchain}.
The same concept has been applied to variants.
For example, the \textit{State channel} pattern involves letting two or more users perform micro-transactions off-chain and regularly storing a hash on-chain to prove the existence of such transactions later on. 
Other studies propose the binding between an off-chain legal contract and an on-chain smart contract, to ensure sensitive data are kept off-chain while only important signatures and states are stored on-chain \cite{23, 13}.

Finally, the third and fourth subcategories are respectively "Wallet and keys pattern" and "Transaction pattern".
They only contain three and one pattern respectively: \textit{Key sharding},  \textit{Hot \& Cold wallet storage}, and \textit{Master \& Sub keys} patterns \cite{13, 14} for the management of block\-chain wallets and keys, as well as the \textit{X-confirmation} pattern \cite{13}.
Although there are only a few patterns in those categories, they have been added as they might contain more patterns later with future studies.

\subsubsection{On-chain Patterns - domain-based Patterns}

The "Domain-based pattern" intermediate category is part of "On-chain pattern", and contains patterns that propose a feature to address a domain-based problem, either for a specific domain or applicable to many.
A list of all the patterns contained in this category is presented in Table \ref{domainpatterns}.

\parindent=0cm
\begin{table}[H]
\begin{tabular}{|P{4cm}|m{8,8cm}|}
 \hline
 \multicolumn{2}{|c|}{\textbf{On-chain patterns - domain-based patterns}} \\
 \hline
 \multicolumn{1}{|c|}{\textbf{\small Subcategory}} & \multicolumn{1}{c|}{\textbf{\small Patterns}}\\
 \hline
    BPM pattern & Block\-chain BP Engine \cite{2} - Smart Contract Activities \cite{2} - Decentralize business process \cite{2} \\
    \hline
    Decentralized identity pattern & Identifier Registry \cite{14} - Multiple Registration \cite{14} - Bound with Social Media \cite{14} - Dual Resolution \cite{14} - Delegate List \cite{14} \\
    \hline
    Big data pattern & Block\-chain Security Pattern for Big Data Ecosystems \cite{17} \\
    \hline
    Multi-domain feature pattern & Block\-chain-based reputation system \cite{2} - Vote \cite{5} - Blocklist \cite{5} - Announcement \cite{5} - Bulletin Board \cite{5} - Randomness \cite{9} - Poll \cite{9} - Selective Content Generation \cite{14} - Time-Constrained Access \cite{14} - One-Off Access \cite{14} - Digital Record \cite{25} - State machine \cite{27} \\
 \hline
\end{tabular}
\caption{On-chain patterns - domain-based patterns}
\label{domainpatterns}
\end{table}

For BPM, 3 patterns have been identified, all proposed in \cite{2}: the \textit{Block\-chain BP Engine} pattern, that enables collaborative business processes by storing and executing a business process through a smart contract, the \textit{Smart contract activities} pattern where business logic activities are stored in a single smart contract for execution, and the \textit{Decentralize business process} pattern that uses block\-chain as a software connector for collaborative business process execution.

Regarding decentralized identity patterns, 5 design patterns have been extracted from \cite{14}.
The first one, \textit{Identifier registry} pattern, proposes the usage of smart contracts to establish a mapping between a DID (Decentralized Identifier), a unique identifier for a human within a domain, and the location of off-chain storage attributes.
Here, the DID is managed using a private key used to prove the ownership of an identifier.
If the key is lost, the \textit{Delegates list} pattern can be used to retrieve this ownership.
To protect user privacy, multiple identifiers can be created using the \textit{Multiple identifiers} pattern.
An identifier can also be mapped to a social media account through the \textit{Block\-chain \& Social Media Account Pair} pattern, to improve the trustworthiness of both social media account and identifier.
Finally, the \textit{Dual resolution} pattern helps to use a DID to enable communication with another entity through its own DID.

One pattern has been identified for the "Big data pattern" category: the \textit{Block\-chain Security Pattern for Big Data Ecosystems} pattern leverages block\-chain to register operations performed on a data store \cite{17}.

The "Multi-domain feature pattern" subcategory groups 12 patterns that propose on-chain features to address problems found in multiple domains.
For example, the \textit{Poll} and the \textit{Vote} patterns \cite{9,5} can be used to take collaborative decisions on-chain, the \textit{Time-constrained access} or the \textit{One-Off Access} patterns \cite{14} let users give access to off-chain resources from an on-chain authorization smart contract, and the \textit{Randomness} pattern \cite{9} can be used to generate random numbers on-chain, a difficult task.

\subsubsection{On-chain Patterns - smart contract Patterns}

The second intermediate category of "On-chain patterns" is the "Smart contract pattern".
In a decentralized application, smart contracts are often the most important pieces.
Many sensitive operations can be performed on them, such as storing and transferring cryptocurrencies.
Therefore, maximal security in smart contract operations is paramount, and well-designed access control functions must be implemented to support it.
Managing them is also difficult, as a smart contract code is immutable once deployed.
Thus, the on-chain smart contract architecture must be adequately designed to tackle the inflexibility of smart contracts and ensure they fill their initial goals while being easily upgradeable if needed.
Finally, they often have to be efficient, as for public block\-chains developers and users have to pay for deploying and executing smart contract functions.
Each of those topics is important for the development of smart contracts and has its own subcategory, presented below.
Table \ref{scpattern} lists all patterns in this category.

\parindent=0cm
\begin{table}[H]
\begin{tabular}{|P{4cm}|m{8,8cm}|}
 \hline
  \multicolumn{2}{|c|}{\textbf{On-chain patterns - smart contracts patterns}} \\
  \hline
 \multicolumn{1}{|c|}{\textbf{\small Subcategory}} & \multicolumn{1}{c|}{\textbf{\small Patterns}}\\
 \hline
    Contract management pattern & Migration \cite{5} - Inter-family communication \cite{6} - Data Contract \cite{11,13,20,27} - Factory Contract \cite{11,13,38,34,37} - Proxy Contract \cite{11,27,34,37,20,38} - Flyweight \cite{11,38,34} - Satellite \cite{27} - Contract Registry \cite{13,27} - Contract Composer \cite{37} - Contract Decorator \cite{37} - Contract Mediator \cite{37} - Contract Observer \cite{37} \\
    \hline
    Contract security pattern & Fork check \cite{9} - Emergency Stop \cite{11,35} - Mutex \cite{11,35} - Contract Balance Limit \cite{11,35} - Automatic Deprecation \cite{27} - Speed Bump \cite{35} - Rate Limit \cite{35} - Check Effect Interaction \cite{11, 35} - Time Constraint \cite{9} - Termination \cite{9} - Math \cite{9} \\
    \hline
    Contract efficiency pattern & Incentive Execution \cite{11,13} - Tight Variable Packing \cite{11} - Limit storage \cite{20} - Minimize on-chain data \cite{20} - Limit external calls \cite{20} - Fewer functions \cite{20} - Use libraries \cite{20} - Short constant strings \cite{20} - Limit modifiers \cite{20} - Avoid redundant operations \cite{20} - Write values \cite{20} - Pull payment \cite{27} - Publisher-Subscriber \cite{34,38} - Challenge Response \cite{36} - Low Contract Footprint \cite{36} \\ 
    \hline
    Contract access-control pattern & Judge \cite{5} - Embedded Permission \cite{5,11,13,9} - Dynamic Binding \cite{11} - Multiple authorization \cite{13,37} - Off-chain secret enabled dynamic authentication \cite{13} - Ownership \cite{27} - Access Restriction \cite{27} - Hash Secret \cite{37} \\
 \hline
\end{tabular}
\label{scpattern}
\caption{On-chain Patterns - smart contracts Patterns}
\end{table}

The first subcategory "Contract management pattern" is about organizing well smart contracts in the decentralized application architecture.
12 different patterns have been labeled with this subcategory.
Some of them address the separation of concerns, between the dApp entry point, features, and data.
The most-frequently mentioned pattern is the \textit{Proxy} pattern \cite{11, 20, 27, 34, 37, 38}.
Usually implemented in traditional software engineering to wrap an object only accessible by it, this pattern is used in block\-chain to wrap a smart contract (the object) into another one (the proxy).
A full description of this pattern is given in a dedicated part of Subsection \ref{p-proxy}.
Another pattern for separation of concerns is the \textit{Data contract} that decouples data from functions in two separate contracts \cite{11, 13, 27}.
The \textit{Flyweight} pattern is similar in functioning, but consists in storing data used by multiple contracts in one place \cite{11,34,38}.
Finally, a mentionable pattern is the \textit{Satellite} that can be used to decouple features that are more likely to change from features that will not change over time \cite{27}.

The second subcategory, "Contract security pattern", is filled with 11 patterns.
Most of them target Solidity-based contracts.
Solidity is a programming language for smart contracts deployed on Ethereum block\-chain networks.
One usage of such patterns is the restriction of access to smart contracts functions when it is needed.
To cite a few of them, the \textit{Termination} pattern consists in locking the contract to prevent any further function call \cite{9}.
It is also possible to use the \textit{Emergency Stop} pattern to simply halt its functioning until reactivated. 
This can be used for instance to protect the contract against abusive withdraw of funds \cite{11,35}.
The \textit{Speed bump} \cite{35}, \textit{Rate Limit} \cite{35}, and \textit{Time constraint} \cite{9} patterns are used to implement time limitations when executing smart contract functions.
Some other patterns aim to protect the correct execution of a function.
The \textit{Check-Effects-Interaction} pattern \cite{35} guarantee a safe execution of the function by first, checking the satisfaction of preconditions, then applying the modifications on the contract, and finally applying modifications on other external contracts, if needed.
Also, some other interesting patterns are the \textit{Mutex} pattern \cite{11, 35} that protects the access to a used resource, or the \textit{Contract Balance Limit} pattern \cite{11, 35} to ensure that the smart contract does not hold too many funds, to mitigate the risk of losing all the funds if compromised.

15 patterns have been added in the third subcategory, "Contract efficiency pattern", and also target Solidity smart contracts.
In Ethereum, a smart contract user must pay a defined amount of Ether, the native cryptocurrency of the network, to deploy or interact with the contract.
The more the function stores data or perform complex operations, the more it will cost the user.
Design patterns in this section help to reduce the fees associated with the deployment, storage, or execution of smart contract functions.
For on-chain storage reduction, many patterns have been proposed in \cite{20}: the \textit{Limit storage} or \textit{Minimize storage data} patterns in general, or \textit{Fewer functions} and \textit{Limit modifiers} to reduce function overhead and code size.
The \textit{Short constant string} pattern can also be used to limit on-chain storage by limiting the size of strings to prevent a high consumption of storage size.
\textit{Tight variable packing} pattern, as proposed in \cite{11}, can also be a solution to reduce storage size by storing data in the smallest unit possible (e.g., Uint8 instead of default Uint256 to store a number below 256).
At computation, the \textit{Avoid redundant operations} and the \textit{Low contract footprint} patterns can help reduce the complexity of operations, thus saving costs \cite{11,36}.
This taxonomy also places in the \textit{Contract efficiency pattern} subcategory patterns that help to keep on-chain data accurate.
For example, the \textit{Incentive execution} pattern \cite{11,13} refunds or reward users that call a specific function to update contract data, as no update can be done by the contract itself without external intervention.

The last subcategory is the "Contract access control pattern" and concerns the permission management of contracts.
9 patterns constitute this category.
The most important one is the \textit{Embedded permission} pattern (also called \textit{Access Control} or \textit{Authorization)}, mentioned by 4 papers \cite{7,9,11,13}, that consists of encoding permission in a smart contract for sensitive functions.
Only authorized addresses will be able to call those functions.
One variant is the \textit{Owner} pattern \cite{27}, which  defines a contract owner as the solely entitled person to execute specific functions.
Authorization to execute a function can also require multiple signatures at the same time.
A pattern named \textit{Multiple Authorization} \cite{13, 14, 37} consists in defining a set of addresses in the contract, where a fraction of them is required to execute a function.
Another noteworthy pattern is the \textit{Judge} pattern \cite{5}, which lets users vote to elect a trusted third party.
The winner is given the authorization to update the smart contract with fresh information, as an \textit{Oracle} could do.

\subsubsection{On-chain Patterns - Data management pattern}

The last intermediate category of "On-chain patterns", "Data management pattern", proposes patterns related to the storage, migration, and encryption of on-chain data.
The complete list of patterns contained in this category is given in Table \ref{datamgpattern}.

\parindent=0cm
\begin{table}[H]
\begin{tabular}{|P{4cm}|m{8,8cm}|}
 \hline
  \multicolumn{2}{|c|}{\textbf{On-chain patterns - data management patterns}} \\
 \hline
 \multicolumn{1}{|c|}{\textbf{\small Subcategory}} & \multicolumn{1}{c|}{\textbf{\small Patterns}}\\
 \hline
    Storage pattern & Transparent Event Log \cite{2} - Key-value store \cite{5} - Address mapping \cite{5} - Event log \cite{20} - Tokenisation \cite{13,9,25,5} \\
    \hline
    Migration pattern & Token burning \cite{15} - Snapshotting \cite{15} - State Aggregation \cite{15} - Node Sync \cite{15} - Establish Genesis \cite{15} - Hard Fork \cite{15} - State Initialization \cite{15} - Exchange Transfer \cite{15} - Transaction Replay \cite{15} - Virtual Machine Emulation \cite{15} - Smart Contract Translation \cite{15} \\
    \hline
    Encryption pattern & Commit and Reveal \cite{11,27} - On-chain encryption \cite{13} \\
 \hline
\end{tabular}
\label{datamgpattern}
\caption{On-chain Patterns - data management Patterns}
\end{table}

Regarding the "Storage pattern" subcategory, 5 have been identified.
The most-proposed one is the \textit{Tokenization} pattern \cite{5, 9, 13, 25}.
Through this design pattern, real-life or complex assets can be encapsulated into a token and exchanged on-chain.
A dedicated part of Subsection \ref{p-proxy} gives a detailed presentation of this pattern.
Other forms of data storage can be mentioned: the \textit{Key-value store} pattern to organize data into a resizable store, accessible with keys, or the \textit{Address mapping} pattern where mapping is established between an address and its associated data \cite{5}.
Finally, some patterns propose to store logs of data into event logs, either in a native block\-chain event log (proposed by some block\-chains, such as Ethereum) \cite{20} or in a smart contract \cite{2}.

Two patterns have been added to the "Encryption pattern" subcategory.
Despite the lack of patterns for this subcategory, it still has been added as many patterns will probably be added to this subcategory in the future, following the advances in on-chain encryption strategies such as homomorphic encryption \cite{liang2020circuit} or zero-knowledge proofs \cite{yang2020zero}.
The \textit{On-chain encryption} pattern \cite{13} helps in protecting sensitive on-chain data through symmetric encryption.
Data can then be stored on-chain and be non-readable by anybody who does not have the encryption key.
The main drawback of this pattern is the key leakage threat because data will remain on-chain forever, even in case of a leak.
The \textit{Commit and Reveal} pattern works differently: some values are kept secret during the commit phase and revealed when needed \cite{11, 27}.
It is possible to attest that the revealed value was the same as the one committed in secret.
Through this pattern, it is possible to commit some data without revealing its content.

In the last subcategory, "Migration pattern", 11 design patterns for data migration are included.
All of those patterns were found in \cite{15}, which proposes a pattern collection for data migration.
To mention a few of them, the \textit{Snapshotting} pattern consists in saving a copy of states, smart contracts, and transactions on the source block\-chain to transfer them to the target block\-chain later.
This operation can be done using the \textit{State initialization} or the \textit{Establish genesis} patterns to respectively transfer states from source to target block\-chain or set states in the first block of target block\-chain (i.e., genesis block).
Besides existing data, the code of useful smart contracts must also be changed to fit the target block\-chain; this can be done using the \textit{Smart contract translation} pattern.

\subsubsection{Architectural Patterns and Idioms}

To conclude this section, other patterns that do not belong to the design pattern taxonomy are introduced.
This sample of patterns contains 14 idioms \cite{11, 20}.
They all concern Solidity, a smart contract programming language for the Ethereum block\-chain, and address smart contract efficiency.
As presented before, users have to pay for smart contract function execution on a public block\-chain.
Proposed idioms help to reduce execution fees in various ways: for example, \textit{Packing variables} or \textit{Packing booleans} patterns can be used to reduce variable required storage with a smart ordering of variables in the code, as background variables are grouped by the compiler in 32-bytes slots.
More efficient structures can be selected to save space, thus costs, using \textit{Uint* vs Uint256} and \textit{Mapping vs Array} patterns.
Ether can also be returned to the user when using the \textit{Freeing storage} pattern, that consists in deleting unused variables or smart contracts.

Additionally, 3 architectural patterns have been identified.
In \cite{24}, authors propose 3 related architectural patterns.
The first one, \textit{Self-generated transactions}, let the responsibility for the user to create and sign transactions to interact with block\-chain smart contracts.
It ensures maximal security, as they keep control of their keys at all times and can verify the code to ensure correct behavior, but it leads to poor user experience and expertise is required.
To facilitate this task, they can use a browser wallet (e.g., Metamask\footnote{https://metamask.io/}) to generate and sign transactions.
Another pattern, \textit{Self-Confirmed Transactions}, is a tradeoff between security and usability as the website is in charge of generating transactions and the user is given the choice of signing them or not, using a browser wallet.
Finally, the \textit{Delegated Transactions} offers the most convenient experience for users, as the website handles all the block\-chain-related operations.
However, trust towards the website is mandatory, as they have full control of keys and wallets.

\subsection{RQ3: What are the most frequently mentioned patterns for each section, and their variants across identified patterns?}

In this subsection, four patterns are introduced in detail, using the Alexandrian form, a pattern format described in the subsection \ref{patternsbg}.
Exploiting the taxonomy, we only selected the most representative patterns in every subcategory, based on the number of references in the corpus of papers. 
Whenever possible, the formalization synthesizes each using the description of the mentioned academic work. 
We completed them with our own analysis of the pattern whenever specific information required by the pattern format was found missing.

\subsubsection{Off-chain Data Storage pattern}
\label{p-offchain}

The \textit{Off-chain data storage pattern} consists in storing a hash of off-chain data in a smart contract, to be able to verify the off-chain data integrity later.
This pattern belongs to the "On/off-chain interaction pattern" category and has been found 6 times in the corpus of papers.

\textbf{Context -} As the block\-chain is replicated among nodes, every node has a copy of it.
Some applications might consider using block\-chain to store a large amount of data, ensuring their integrity \cite{11, 13}.

\textbf{Problem -} Allowing users to store on-chain data without any limit of storage could hamper the network functioning.
Therefore, many block\-chain networks enforce a block size limit to limit the size growth of block\-chain over time.
Even if the size limit suits the needs of the user, storing data on-chain is prohibitively expensive.
Thus, how can the user store data on-chain while taking advantage of block\-chain immutability and integrity \cite{13}?

\textbf{Forces -} Using this pattern implies balance forces.
The first one is cost, as storing data on-chain is expensive and even more if using a smart contract to keep the possibility to perform operations on them directly on-chain.
Then, scalability, because storing large files on a block\-chain is difficult as they are replicated across all nodes \cite{13}.
Finally, immutability level has to be considered: storing a hash on-chain does not offer the same protection as storing the file itself.
Indeed, it can still be modified or deleted off-chain.

\textbf{Solution -} Store the data off-chain, then calculate a hash of those data.
Store the result on-chain in a smart contract, possibly associated with metadata (e.g., resource location, description, ...) \cite{11,36}.
As hashing data is a one-way function, data confidentiality is preserved, and users can check the integrity of their data using the immutable hash stored on-chain \cite{13}.

\textbf{Example - } A company that wants to store proof that a legal contract is signed can hash the contract after its signature and store the result on-chain.
Thus, if another company denies the authenticity of a contract, it is possible to prove the existence of the document as well as its metadata (e.g., signature time).

\textbf{Resulting context -} Data are kept off-chain, and stay confidential, but their integrity can still be accessed using the on-chain hash.
It is inexpensive to store the hash on-chain compared to the file itself, considering the size of such file is large.
However, the file is still vulnerable to deletion or tampering, as the hash itself cannot help retrieve a lost file or deleted content.
Adequate measures must be taken to preserve off-chain data.

\textbf{Related patterns -} According to \cite{14}, this pattern is directly related to the \textit{Low contract footprint pattern} in \cite{36}, as the latter propose to minimize the number and size of on-chain transactions to save costs, notably with optimizing write operations.
As the \textit{Off-chain data storage pattern} only stores a hash on-chain, this cost is kept low.

\textbf{Known uses -} The Government of Estonia’s e-health solution utilizes block\-chain as a "fingerprint" registry to ensure the integrity of e-health records \cite{25}.
Factom\footnote{https://www.factom.com/}, a block\-chain for building records systems, implements this pattern by systematically hashing files sent to the block\-chain.
Only the hash is kept on-chain after the operation.

\subsubsection{State Machine pattern}
\label{p-statemachine}

The \textit{State machine pattern} proposes to manage smart contract state transitions through state machines, to break the problem of state changes into simple state transitions.
It belongs to the "Domain-based pattern" intermediate category.
As each of the patterns included in this category has only been found one time in selected papers, we decided to select the \textit{State machine pattern} for a thorough introduction as this pattern can be used in many different scenarios, including basic implementations of smart contracts.

\textbf{Context -} When leveraging smart contracts, state changes are often performed.
Depending on the purpose of the smart contract, many states changes might occur during its lifecycle.

\textbf{Problem -} A smart contract might be difficult to design if many state changes occur, as complex logic must be implemented.

\textbf{Forces -} Some forces are bound to the usage of this pattern: the complexity of the smart contract to design and its efficiency, as depending on the implementation of the state changes part, the contract might be efficient or cumbersome to use.

\textbf{Solution -} Apply a state machine to model and represent different contract stages and their transitions in the smart contract \cite{27}.

\textbf{Example -} A company that wants to leverage a business process on-chain with multiple steps that might trigger automatic operations might be tempted to use the \textit{State machine pattern} in order to model and perform the state changes within the contract.

\textbf{Resulting context -} The state machine breaks complex problems into simple states and state transitions \cite{27}, resulting in a more efficient smart contract.

\textbf{Related patterns -} In the \textit{Confidential and pseudo-anonymous contract enforcement pattern} \cite{23}, a state machine can be employed in the smart contract used by the pattern to handle state changes of the associated legal contract on-chain.

\textbf{Known uses -} The DutchMachine smart contract implements a state machine for handling auctions \cite{27}. 

\subsubsection{Tokenization pattern}
\label{p-tokenization}

The third presented pattern is the \textit{Tokenization pattern}.
Classified in the "Data management pattern" intermediate category and mentioned 4 times, this pattern consists in representing an asset by a token, to facilitate its exchange on block\-chain networks.

\textbf{Context -} Through a block\-chain network, it is possible to send transactions and interact with smart contracts without any third party as an intermediate.
Such a network enables the exchange of value directly between one user to another, notably with the exchange of native cryptocurrency.

\textbf{Problem -} Native fungible block\-chain tokens (e.g., Bitcoin, Ether) often serve as the native cryptocurrency of the associated block\-chain network.
In some cases, they can also be used as token support to track assets, but their capabilities are limited.
Indeed, extending the concept of value exchange for other types of assets (e.g., other currencies, art, houses, ...) is not a straightforward process due to the dissimilarity between those assets.

\textbf{Forces -} Some forces are bound to this pattern: authority, as it must be ensured that the on-chain asset is the authority source of the correlated asset \cite{13}, and liquidity, as block\-chain can enable a frictionless exchange of value.

\textbf{Solution -} Model many types of assets on block\-chain using tokens.
Two types of tokens can be differentiated: fungible tokens that are indistinguishable from each other, and non-fungible tokens (NFTs), representing a unique asset with its own properties.
Smart contracts can thus be used as a data structure to handle the tokens and associated operations (transfer, deletion, ...) \cite{13}, but also enhance their capabilities.

To illustrate, Ethereum proposes two different standards to create fungible and non-fungible tokens using smart contracts, that are respectively ERC20\footnote{https://ethereum.org/en/developers/docs/standards/tokens/erc-20/} and ERC721\footnote{https://eips.ethereum.org/EIPS/eip-721} tokens. 
Using these standards simplifies the usage of tokens, as on-chain applications and users can rely on standard interfaces to interact with all of the smart contracts that implement tokens for their usage.
Other standards exist in the Ethereum ecosystem to improve their usability in different contexts.
For instance, the ERC1155 can also be mentioned as it allows the usage of both fungible and non-fungible tokens (ERC20 and ERC721) in the same smart contract.
ERC998-based tokens go even further by regrouping multiple tokens under a single token (commonly called a basket).
This simplifies their exchange between users and enables other use cases (e.g. a service proposing users to invest in a specific basket of tokens all-at-once).
A variant, the ERC3664, allows the combination of multiple NFTs to a single one.
This composability of NFTs is notably useful in the gaming industry (e.g. a set of items merged into a better one).

Where tokens can be used to represent different types of assets, they can also be used for other purposes.
One of the most popular uses in this context is token governance: depending on their amount of owned tokens, users could vote on important decisions.
For instance, owning governance tokens that represent a share of an on-chain fund, users could vote about the usage of those funds, such as their investment in other protocols.
Another similar concept is staking, notably for Proof-of-Stake block\-chains: by locking a defined amount of their tokens at stake, users could be entitled by the consensus algorithm to create new blocks.

\textbf{Example - } A real estate company can use non-fungible tokens to represent the ownership of houses directly into the block\-chain. 
Ownership of a house can then be directly exchanged on-chain, and a complete history of transactions can be retraced for a house.

\textbf{Resulting context -} Assets are tokenized on-chain and can be easily sent between users.
Using smart contracts, many features can be implemented along with the tokens, such as royalties, sales, or burn (i.e., destroying tokens).

\textbf{Related patterns -} The \textit{Address mapping pattern} can be used as a complement to map block\-chain accounts (e.g., public addresses) with owned tokens.
The \textit{Poll} pattern might use the \textit{Token} pattern to materialize votes as tokens and keep track of them.

\textbf{Known uses -} The \textit{Tokenization} pattern has already been applied in a tremendous number of domains.
For instance, stablecoins (e.g., Tether\footnote{https://tether.to/}), consist in emitting fungible tokens on-chain that keep the same value as an underlying asset (e.g., US Dollar) using different strategies. 
This enables many other use cases relying on the usage of fiat currencies, such as frictionless currency swap.
Another use case is the usage of NFTs in art.
Many artists have digitalized their art as NFTs to sell it on on-chain marketplaces, such as OpenSea\footnote{https://opensea.io/}.

\subsubsection{Proxy contract pattern}
\label{p-proxy}

The fourth and last presented pattern is the \textit{Proxy contract pattern}.
It belongs to the "Smart contract pattern" intermediate category and appeared 6 times in found patterns.

\textbf{Context -} In a block\-chain, data becomes immutable after addition.
This concept also applied to smart contracts, that cannot be modified after their deployment on-chain \cite{20}.

\textbf{Problem -} If a smart contract must be changed, for diverse reasons (upgrades, bug correction, ...), the developer has to deploy another version of the contract and manually change the other contracts that reference the old contract \cite{20}.
In the best case, this is a cumbersome task, and it might even not be possible in certain cases.

\textbf{Forces -} The problem requires balancing the following forces: first, immutability, as deployed smart contracts are designed to be immutable, and upgradeable, as proposing features to allow upgradeability enhances designed smart contracts.

\textbf{Solution -} Using a proxy contract, a user can query the latest version of a target contract.
The proxy contract will relay the request to the target contract \cite{27}.
By replacing the reference of the target contract with a new one, it is possible to easily upgrade parts of the decentralized application \cite{20}.

\textbf{Example -} A user can request a proxy contract as the bridge for a decentralized application, such as the latest version of a decentralized cryptocurrency exchange.

\textbf{Resulting context -} Proxy contracts can be used to easily access the latest version of a contract, without requiring storing the latest contract addresses off-chain.
Reference updates can easily be performed by requesting the proxy contract with the latest contract address.

\textbf{Related patterns -} The \textit{Data contract pattern} can be implemented along the \textit{Proxy contract pattern} as the proxy will allow updating the logic used to access the data contract without updating the data contract itself.
The \textit{Contract registry pattern} is related to the \textit{Proxy contract pattern}, as the contract registry has a reference to all the latest versions of the contracts, where the proxy only references one contract.

\textbf{Known uses -} A security company named OpenZeppelin proposes a generic implementation of the \textit{Proxy contract pattern} for Solidity-based smart contracts\footnote{https://blog.openzeppelin.com/proxy-patterns/}.
Uniswap, a decentralized exchange on Ethereum, uses proxy contracts to forward user transactions to the exchange smart contract\footnote{https://etherscan.io/address/0x09cabec1ead1c0ba254b09efb3ee13841712be14}.

\subsection{RQ4: Are some of the patterns equivalent to existing software patterns?}
Since the first collection of design patterns released by the GoF \cite{gamma1995elements}, many patterns have been proposed that can be applied in many contexts.
As dApps have many similarities with traditional applications, one aspect this study investigates are the links between existing software patterns and proposed block\-chain-based patterns, either through the creation of variants or the direct usage of existing patterns in block\-chain applications.
Table \ref{softwarepattern} introduces the list of all identified software patterns.
It has been found that 22 extracted patterns mention references to existing software patterns, where 16 of them directly arise from the GoF design pattern collection.
A possible explanation is that smart contracts have many similarities with objects, many GoF design patterns can be applied to them.
For example, the \textit{Factory} pattern is used to create instances of smart contracts from a factory contract, as it can be used in OOP (Object-Oriented Programming) to create objects from one.
On top of that, using GoF patterns with smart contracts can help to tackle their lack of flexibility, a difficult aspect to manage in dApp development.
To illustrate, where the \textit{Proxy} pattern is a good practice for protecting the access of sensitive objects, it is even stronger with smart contracts, as it can dynamically change its target from one contract to another, a mechanism used to upgrade an old smart contract by switching to the new one.

\begin{table}[h]
\begin{tabularx}{\textwidth}{|X|X|}
\hline
\multicolumn{1}{|c|}{\bfseries GoF pattern} & \multicolumn{1}{c|}{\bfseries Other pattern} \\ \hline
Proxy (4) - Factory (3) - Flyweight (3) - Chain of responsibility (2) - Observer (1) - Facade (1) - Memento (1) - Composite (1)
&
Mutex (2) - Publisher-Subscriber (2) - Snapshot (1) - Layered design (1)
\\
\hline
\end{tabularx}
\caption{Existing software patterns mentioned by selected studies and their occurrences.}
\label{softwarepattern}
\end{table}

\subsection{RQ5: What are the applications of identified patterns?}

Looking at the domain applications, 7 papers out of 20 targeted a specific domain, such as healthcare, big data, decentralized identity, record management, financial services, and BPM.
The proximity between some of the patterns and their application domain is the reason they have been classified in the \textit{Domain-based pattern} subsection of the taxonomy.
In \cite{2}, specific patterns for BPM have been proposed.
They might be applied in other solutions, but their main purpose is bound to business process management.
In other cases, some patterns are presented as a domain-agnostic solution coupled with implementation details in a specific application domain.
For instance, \cite{34} proposes an adaptation of GoF patterns to serve healthcare solutions, using block\-chain.

From a technological standpoint, 6 of the 20 selected papers propose patterns for specific block\-chain technology.
In those papers, 5 are focusing on Ethereum, and more specifically Solidity smart contracts.
Indeed, a growing interest is shown by academics and businesses for Ethereum since its release in 2016, as its mainnet is currently the most-adopted public block\-chain network for smart contract development.
In this context, software patterns support many aspects of Solidity-based smart contracts.
As seen before, found patterns mainly address the efficiency and the security of smart contracts, two major aspects to consider when developing Solidity-based decentralized applications.
Another paper introduced a pattern for the Hyperledger ecosystem, more specifically for Sawtooth, a modular block\-chain technology\footnote{https://www.hyperledger.org/use/distributed-ledgers}.
Looking at patterns themselves, over the 160 non-unique patterns retrieved, 28 of them were not mentioning the usage of smart contracts, 79 of them mentions the usage of smart contracts without any precision on used technology in the pattern \textbf{Solution}, and 53 patterns are proposed in the context of using a specific technology (e.g., Ethereum).
However, we found that some of the patterns might be proposed in a more generic form, thus allowing its application to other technologies.
This might be the ground for future research in this domain.

\subsection{RQ6: What are the current gaps in research on block\-chain-based patterns?}

Regarding current gaps in research on block\-chain-based patterns, the lack of non-design patterns can be mentioned.
Among the 120 patterns retrieved, only 3 of them are architectural patterns and 14 of them are idioms.
Although design patterns are a very compelling solution for the design of robust and efficient applications, exploring new forms of block\-chain architectures, then formalizing them as architectural patterns could benefit a lot to block\-chain dApp design.
Taking back the examples mentioned in Subsection \ref{rq2}, \cite{7} shows the strong impact on software quality using architectural patterns.
On one side, applying the \textit{Self-Generated Transactions} pattern means letting the task of signing transactions to users on the client-side, thus ensuring no one aside the client has access to the keys.
On the other side, using the \textit{Delegated Transactions} pattern lets full control of the funds to the application.
This can be convenient for users without knowledge of using a block\-chain wallet but adds a potentially vulnerable third party into the balance.
Such research could be conducted by exploring the existing literature or applications to find innovative ways of organizing decentralized application components.
For example, \cite{tonelli2019implementing} proposes a microservices system where smart contracts are services themselves.
As the \textit{Microservices} architectural pattern already exists, adapting it for block\-chain could lead to a new way of designing a loosely coupled smart contract system with its own advantages and liabilities.

Regarding the idioms, and the other smart contract patterns found, all of them deal with Solidity, except one (Hyperledger Sawtooth). 
Although Ethereum is the most used public block\-chain for decentralized applications as of today, yet other languages could be considered.
Rust, a high-level compiled language, is used for smart contract development by many block\-chain technologies, such as ink! from Polkadot\footnote{https://github.com/paritytech/ink}, or Rust for Solana\footnote{https://github.com/solana-labs/solana}.
Formalizing new idioms and patterns in this context could help improve code quality and security.
In addition, existing patterns in Solidity could also be translated for other block\-chains.
As an example, the \textit{Freeing storage} idiom from \cite{20} could also be applied to other public block\-chains where freeing the storage refunds a defined amount of money.

\section{Threats to Validity}
\label{threats}

In this literature review, the Kitchenham et al. methodology has been applied to systematically conduct the study, from the selection of papers to the collection of data \cite{Kitchenham07guidelinesfor}.
Although using this method helps to limit the bias in our study, some internal threats can appear due to manual steps performed.
Regarding the query used, applying it to titles, abstracts, and keywords improve the precision of the request, yet some papers might have been missed.
To overcome this, backward and forward snowballing has been used to retrieve papers that cited or have cited studies found while performing the systematic literature review.
The categorization and grouping of patterns can also lead to manual errors, as this is performed mainly from data collected from the patterns, that might lack accuracy.
In parallel, great attention has been paid to not merging patterns that are not strictly identical, to avoid missing variants of patterns that serve in different contexts.

The taxonomy construction is also subject to bias.
For instance, even if different methods were used to generate category names, the final decision is up to the taxonomy builders.
Selecting the high-level dimensions (meta-characteristics) is also a subjective task that has a high impact on the construction of the taxonomy.
To limit such bias, the methodology from Nickerson et al. was applied \cite{nickerson2013method}.
Also, found patterns have easily been classified into the final version of the taxonomy, hence, the produced version of the taxonomy satisfies the goals initially described before its construction.


\section{Related Works}
\label{related}

Using systematic literature reviews to collect patterns is a strategy that has already been used in other fields.
For instance, \cite{juziuk2014design} has gathered 206 design patterns on multi-agent systems (MAS) from the literature.
Authors have also identified the links between found patterns and proposed classification to group patterns under different categories and subcategories.
The study also mentions several research gaps in the literature for MAS, such as the lack of standardization when describing a pattern despite the existence of several pattern formats, the lack of links between patterns that do not belong to the same category, and the lack of mentioned applications of presented patterns.
Our study also shares the same conclusions.
In \cite{osses2018exploratory}, 44 architectural patterns are extracted from a corpus of 8 papers about microservices.
A taxonomy is also provided to classify the different patterns.
It has been found that identified patterns are mostly bound to five quality attributes: scalability, flexibility, testability, performance, and elasticity.
\cite{washizaki2020landscape} has also performed a systematic literature review of IoT software patterns, and has collected 143 architecture and design patterns from a corpus of 32 papers.
They have also identified that 57\% of all found patterns are non-IoT patterns, thus meaning IoT systems are designed through a conventional architecture perspective, something that we also identified in this study through the "On/off-chain interaction pattern" category as well as GoF-based design patterns.
Our study follows the same path as others by proposing a taxonomy and a collection of patterns.
To the best of our knowledge, this is the first attempt in the block\-chain-based pattern literature to propose such work.


\section{Conclusion and Future Works}
\label{conclusion}

Ensuring the high quality and efficiency of newly built decentralized applications is a challenge of uttermost importance for the future of block\-chain.
Software patterns are a promising solution to address this challenge, as they ensure commonly occurring problems in a given context are addressed with extensively tested solutions.
In this paper, a systematic literature review is performed on the available block\-chain pattern literature to identify existing software patterns and classify them into a comprehensive taxonomy.
20 studies have been selected for reading and kept accepted afterward, from where 160 patterns have been extracted. After duplicate removal, 120 unique patterns have been found and regrouped in a taxonomy.
The taxonomy consists of 4 main categories and 15 subcategories and has been built using a construction taxonomy methodology \cite{nickerson2013method}.
This paper also discusses the links between found patterns, but also their relation with existing software patterns such as the GoF design pattern collection.
One finding is that many patterns from this collection are translated into block\-chain patterns, such as the \textit{Proxy} or \textit{Factory} pattern.
Future research could be performed on the translation of GoF patterns that have not been translated yet as block\-chain-based patterns.
It has also been found that while some patterns are described in a very generic form, some variants propose specific forms of patterns based on them, such as the \textit{Oracle} pattern that was derived into 4 different variants in \cite{39}.
Application domains of patterns have also been discussed: among the corpus of papers, we found 3 papers directly linked to domain-based patterns, respectively healthcare \cite{38}, collaborative business processes \cite{2}, and decentralized identity \cite{14}.
Finally, research gaps are addressed: we enlight the scarcity of architectural patterns and idioms for the design of block\-chain-based architectures, and the concentration of patterns on one block\-chain protocol (Ethereum).
Further research in creating architectural patterns and idioms for various block\-chain protocols could benefit the development of robust block\-chain-based applications.

This study is part of a bigger project that aims to empower software architects with a framework for the design and implementation of block\-chain-based applications \cite{six2021decision}, notably with the construction of BLADE (Block\-chain Automated Decision Engine), a decision-making tool to select the most suitable block\-chain and patterns for a given context \cite{10.21494/ISTE.OP.2021.0604}.
The taxonomy proposed in this paper paves the way for the construction of an ontology in a further study, to perform semantic queries (SPARQL\footnote{https://www.w3.org/TR/rdf-sparql-query/}) on its entities (i.e., patterns), and forms the knowledge base of BLADE. 
Building this ontology will also be an opportunity to further explore links between patterns and groups of patterns, an aspect partially left from outside the scope of this paper.
In this paper, patterns were collected from existing studies, yet future works will also consist in extending the current knowledge with new patterns, systematically derived from existing block\-chain-based applications through a literature review, but also by proposing new patterns (i.e., \cite{six2020blockchain}) to enable the design of better-decentralized solutions.
The collected patterns will also be reused through BLADE, as it will recommend patterns from this knowledge set.
Finally, this study also contributes to the state of the art of block\-chain-based patterns, through a taxonomy that will help to classify newly created patterns in comprehensive categories, a systematic literature review to map and describe the existing literature on block\-chain-based patterns within the taxonomy, and highlight research gaps that could be addressed in further studies.

\section*{Declaration of competing interest}

The authors declare that they have no known competing financial interests or personal relationships that could have appeared to influence the work reported in this paper.


\bibliographystyle{elsarticle-num} 
\bibliography{biblio}






\end{document}